\documentstyle [12pt] {article}
\begin{document}
\textwidth 18.cm
\textheight 23.0cm
\topmargin -0.5in
\baselineskip 16pt
\parskip 18pt
\parindent 30pt
\title{ \large \bf The origin of spacetime topology and generalizations
of quantum field theory}
\author{Ulvi Yurtsever \\
{}~~~~~~~~~~~~ \\
Department of Physics \\
University of California \\
Santa Barbara, CA 93106 }
\date{October, 1992 \\
UCSBTH-92-45 }
\pagestyle{empty}
\baselineskip 24pt
\maketitle
\vspace{.2in}

\baselineskip 12pt
\begin{abstract}
\noindent The research effort reported in this paper is directed, in a broad
sense, towards understanding the small-scale structure of spacetime.
The fundamental question that guides our discussion
is ``what is the physical content of spacetime topology?"
In classical physics, this question has a natural and simple answer:
spacetime (as a topological
space) is a bookkeeping device that we invent to
make the description of classical fields (the
observables) easier. More precisely, if spacetime, $(X, \tau )$,
has sufficiently regular topology,
and if sufficiently many fields exist to allow
us to observe all continuous functions
on $X$, then this collection of continuous functions
uniquely determines both the set of points
$X$ and the topology $\tau$ on it. Naturally, however,
this answer does not yield any clues into why
our spacetime is observed to have a very special
(i.e. smooth manifold) topology down to the
smallest scales we can probe, or into whether this
smooth manifold structure persists indefinitely,
at length scales smaller than the smallest observed so far.
To explore these queries, we are led to consider the
original question in the context
of quantum, rather than classical field theory. After all,
in the real world physical fields
(the observables) are not classical (continuous functions)
but quantum operators, and the fundamental
observable is not the collection of all continuous
functions but the local algebra of quantum field operators.
Presently the only examples of local quantum
field algebras that we know how to construct
rigorously (apart from some two-dimensional models)
are the operator algebras of free (linear)
quantum fields propagating on a smooth, globally hyperbolic spacetime.
Since this class of examples is too small,
we find it necessary to generalize
the algebraic notion of ``quantum field" in such a way
that it becomes possible to talk about quantum field theory on
an arbitrary (not necessarily smooth) topological
space on which no notion of spacetime metric
exists a priori. [One interesting offshoot of this
generalization is an algebraic framework for linear
quantum field theory on non-globally-hyperbolic
spacetimes (e.g. spacetimes with naked singularities or
closed timelike curves),
which is the subject of a separate paper submitted elsewhere.]
In pursuing the original
problem further, we develop a still wider
generalization of quantum field theory; this ultimate generalization dispenses
with the fixed background topological space altogether
and proposes that the fundamental observable
should be taken as a lattice (or more specifically a ``frame," in
the sense of set theory) of closed subalgebras
of an abstract $C^{\ast}$ algebra.
Our discussion concludes with the definition and some elementary
properties of these ``quantum lattices" and ``quantum frames."
\vspace{0.5cm}

\end{abstract}

\newpage
\pagestyle{empty}
\baselineskip 16pt
\parskip 16pt
\pagestyle{plain}
\pagenumbering{arabic}
{}~~~~~~

{\bf \noindent 1. Introduction}

Spacetime, general relativity teaches us, is not a fixed, pre-determined
background on which physical processes take place, but a dynamical
entity itself. In classical (relativistic) physics,
the dynamical content of spacetime is primarily geometric, with
the metric and matter fields tied together as
described by Einstein field equations.
In quantum physics, it is widely believed that not only the
geometry of spacetime but also its topology
must be subject to dynamical fluctuations, governed
by an as yet unknown quantum theory of gravity. The scale at which these
fluctuations become significant is set by the Planck length ($l_{P} \sim
10^{-33} \mbox{cm}$). The small
scale structure of spacetime at or below the
Planck length is bound to be very different from that of a
locally Euclidean, four-dimensional
Lorentzian manifold, the structure currently
observed to be accurate down to at least about $10^{-16}
\mbox{cm}$.

The imminent breakdown of the macroscopic structure of
spacetime at sufficiently small length scales (too small to be
accessible by experiments in the foreseeble future)
forces us to take one of two positions: Either the basic manifold structure
(but not necessarily the four-dimensional Lorentzian geometry) of
spacetime is
fundamental and accurate down to arbitrarily small scales, or, the
structure of spacetime below some small length scale ($\geq l_{P}$) is
fundamentally different from that of a point set with manifold
topology. If the first position is adopted, there is in principle no reason
to expect that our basic notions about physical space
will have to be revised radically
in order to understand the quantum theory of spacetime. The standard,
differential-geometry-based approach to classical field theory, along
with the universally accepted principles of quantum mechanics, have
to be capable, eventually, of achieving this ultimate quantization.
This attitude is implicit in much
of the current work on quantum gravity, in
both string theory and the canonical quantization approach.

By contrast, if the second position is adopted, then one cannot escape
the conclusion that understanding the full quantized structure of
spacetime will have to involve extensive and radical changes in our
basic notions about space, field theory and gravity. The argument for
this conclusion can be stated
briefly as follows: In every known example of quantum field theory on a fixed
(manifold) background, the equations of
motion for the classical fields continue to hold after quantization.
Although the expectation values of the fields do not, in general, satisfy the
classical field equations, the field operators themselves (constructed as
operator-valued generalized functions on the background) always do.
(We are assuming, of course, that the nonlinear interaction terms involving
the fields are suitably regularized to make proper sense of the products of
field operators. The difficulties associated with this step are
notorious; however, these difficulties are technical, not fundamental.) In
other words, in standard quantum field theory the equations of motion that
form the starting point for quantization are fundamentally valid,
not only classically but also in the quantum theory. If, on the other hand,
our goal is to quantize spacetime itself (or, what amounts to the same
thing, to quantize gravity), and if spacetime does not have manifold
structure at small-enough length scales, then the classical field
equations, defined as they are
in terms of the macroscopic, manifold ``limit" of
spacetime, have
absolutely no chance of being fundamental: at the quantum
level not even the formulation of these equations makes sense. The
straightforward quantization of any classical field theory (whether
that theory is general
relativity, string theory, or some other extension of
general relativity) cannot bring us a final
understanding of the quantum structure of spacetime. (For a poignant
comparison, consider the understanding one would gain about the small-scale
structure of a fluid from the
quantization of the Navier-Stokes equations.) In short, if we take
the position that the fundamental, small-scale structure of spacetime
is {\it not} that of a manifold, then no
formalism based on differential geometry can be the correct quantum
theory of gravity.

In a series of papers, of which this is the first one, we will adopt this
second position and explore some of its consequences from a specific, and
hopefully novel, point of view. There are other, previous studies
of spacetime structure which adopted a similar attitude towards the
ultimate nature of quantum gravity. Perhaps the best-known among these are
the spin-networks idea of Penrose ([1]), the work on causal sets by
Sorkin and others ([2]), and the more recent work of Isham ([3]) on quantum
topology. Our approach differs fundamentally from these earlier
studies in one respect: We will {\it not}
develop, at the outset, a specific,
``discrete" structure, postulate it to be the fundamental small-scale
structure of spacetime, and proceed to discuss its consequences. Instead we
ask: what is the essential content of our notion of ``space" in physics; or,
in more precise terms, what is the physical origin of spacetime topology?
In essence our answer is simple: Spacetime is not observable
directly; only fields are. Spacetime, then, is a ``bookkeeping"
device, one of indispensible convenience for describing the dynamics
of the physical fields we observe, but not as fundamental as the
fields themselves. This is the guiding principle of our approach, and
it will frame our discussion throughout this and the later papers in the
series. In this first paper, we will mainly develop the foundations for our
viewpoint, i.e. explain how
spacetime structure can be regarded naturally as
``derived" from field theory.
We start with the case of classical fields.

{\bf \noindent 2. The origin of spacetime topology in classical physics}

The most general proper context in which the notion
of spacetime structure can be placed in classical field theory is the
category of topological spaces and continuous maps. By this we mean that
spacetime is a set, $X$, endowed with a topology, $\tau$, and fields are
continuous functions from $(X, \tau )$ into other, target topological
spaces. By contrast, in quantum theory the most general category
for spacetime structure is in fact bigger than the topological
category as we will see shortly.

To render our discussion here more transparent we will make two simplifying
assumptions: First we assume that there exist sufficiently many physical
fields on spacetime to allow us to observe the
family, ${\cal F}$, of all continuous real-valued
functions on
$(X, \tau )$. [Clearly, ``observable" functions include not only the
fields themselves,
but also a large class of (e.g. piecewise analytic) functions
of the (components of) fields.] Next, we will assume that it is possible to
distinguish by observation whether a given collection of real numbers,
$\{ c_{f} | f \in {\cal F} \}$, corresponds to the values assumed by the
observables in ${\cal F}$ at a point of $X$; i.e.,
we assume that given $\{ c_{f} \}$ we can decide whether there exists a
point $x \in X$ such that $c_{f} = f(x) \; \; \forall f \in {\cal F}$.
More sophisticated approaches that do away with these assumptions are
possible; however, we do not expect that our main conclusion will be altered
by relaxing the above assumptions.

The collection ${\cal F}$ is the fundamental observable associated with
the spacetime $(X, \tau )$. How can we reconstruct $(X, \tau )$ from
the knowledge of ${\cal F}$? Obviously, we need to recover both the
point-set $X$ and the topology $\tau$ on it. This can be done
if and only if the following two conditions are satisfied:

\noindent (i): The collection ${\cal F}$ separates points; that is, for
any $x, \; y \in X, \; x \neq y$, there exists a continuous function
$f$ in ${\cal F}$ such that $f(x) \neq f(y)$.

\noindent (ii): The topology $\tau$ is the weakest topology on $X$ that
makes every $f \in {\cal F}$ continuous; in other words, $\tau$ is
generated by subsets of $X$ of the form $f^{-1} (J)$, where $f \in {\cal
F}$, and $J \subset {\bf R}$ is open in ${\bf R}$.

\noindent In other words, if $(X, \tau )$ satisfies conditions (i)
and (ii), then the knowledge of the set of all
classical observables, ${\cal F}$, is equivalent to the knowledge of
both $X$ and $\tau$. In fact, this is precisely the way we
``deduce" the topology of physical space in classical physics: all
neighborhood relationships are determined relative to the measurements
of fields, and fields are declared to be continuous functions at the
outset.

Which topological spaces satisfy the conditions (i)$-$(ii)? To answer
this question, it is convenient to recast the conditions
(i) and (ii) in a slightly
more elegant form as follows: By composing each continuous function in
${\cal F}$ with a fixed homeomorphism of ${\bf R}$ onto the
unit interval
$(0,1)$, we can, without loss of generality, replace ${\cal F}$
with the collection of all (bounded) continuous functions $X
\longrightarrow [0,1] $ (which collection we will still denote by ${\cal
F}$). Let $I$ denote the closed interval $[0,1]$, and let
$I_{\cal F} \equiv \prod_{f \in {\cal F}} I$,
the Cartesian product of ${\cal F}$-many copies of $I$, with one copy for each
distinct $f \in {\cal F}$. $I_{\cal F}$ is
a compact Hausdorff space under the natural product topology.
We can now construct the following canonical map
\begin{eqnarray}
i & : & X \longrightarrow I_{\cal F} \; , \nonumber \\
i & : & x \mapsto i(x) \in I_{\cal F} \; \mbox{where} \;
[i(x)]_{f} = f(x) \; \; \forall f \in {\cal F} \; .
\end{eqnarray}
It is easy to see that $i$ is continuous in general, and that the
conditions (i)$-$(ii) are equivalent to the statement that $i$ is an
imbedding, i.e. a homeomorphism onto its image $i(X) \subset I_{\cal
F}$. Therefore, in Eq.\,(1) we have an explicit picture of the
reconstruction of $(X, \tau )$ from the observed data ${\cal F}$ (and
the subset $i(X) \subset I_{\cal F}$, which is distinguishable in
$I_{\cal F}$ by our assumption). The question we posed at the beginning
of the paragraph now becomes: for which spaces $(X, \tau )$ the map $i$
given by Eq.\,(1) is an imbedding? The answer is well known in topology,
and these are precisely the so-called Tychonoff spaces, i.e. Hausdorff
spaces $X$ with the property that for any closed subset $A$ and a point
$x$ not in $A$ there exists a continuous function $f : X \longrightarrow
{\bf R}$ such that $f(A)=a$ and $f(x) = b$, where $a \neq b$ (see [4],
Sect.\,14). Tychonoff spaces are regular (in fact $T_{3}$ since $T_{2}$
holds by definition), but not every $T_{3}$ space is Tychonoff.
Subspaces and products of Tychonoff spaces are Tychonoff, and every locally
compact Hausdorff space (hence every manifold)
and all metric spaces are Tychonoff. An interesting property is that a
{\it connected} Tychonoff space is
either a trivial, one-point space, or contains
uncountably many points.

We have thus demonstrated that in classical field theory spacetime
topology can be recovered completely from a knowledge of the fields
(continuous functions) provided the background, $(X, \tau )$, is
a Tychonoff space. This answer is unsatisfactory
for a number of reasons: First
of all, it does not bring any insight into why our spacetime is observed,
at least down to the smallest length
scales we can probe, to have a smooth manifold structure. Although
Tychonoff
spaces and manifolds have in common the property that any nontrivial,
connected subspace is uncountable, there remains a large ``evolutionary"
distance between a typical Tychonoff space and a manifold. Indeed,
general Tychonoff spaces are not even metrizable. Our answer is
also unsatisfactory in that it does not provide an interesting
alternative to manifold structure that might serve as a candidate
for the small-scale structure of spacetime. Tychonoff spaces are more
general than manifolds, but they hardly seem promising as candidates
for the fundamental structure of spacetime at arbitrarily small length scales.

These deficiencies, however, are expected, or should have been expected,
because the above questions which our approach so far failed to answer
are quantum-mechanical in nature, not classical. After all, the
expectation that spacetime at small-enough length scales is likely to
have non-manifold structure has a quantum-mechanical motivation. Also,
in the real world physical fields (the observables) are not classical,
continuous
functions but operator-valued quantum fields. We have
to formulate our approach in the context of quantum field theory to
be able to address the fundamental issues relating to small-scale spacetime
structure.

We have argued that the only observable associated with topological
structure in classical field theory is the collection ${\cal F}$ of all
continuous, real-valued functions on spacetime.
The main challenge that we will face for the rest of this paper is the
formulation of a suitable analogoue of the observable ${\cal F}$ in
quantum field theory. The best (and perhaps the only) approach
to quantum field theory in which this formulation
can be worked-out is the algebraic approach. For reviews of the
essential aspects of the algebraic approach to quantum field theory in
curved spacetime, we recommend the reader consult Refs.\,[5] and [6];
for an extremely brief overview see
Sect.\,1 of [7].

Recall that the fundamental construction in the
algebraic approach is the local algebra of field operators, ${\cal A}$,
along with the ``net" of closed subalgebras, $\{ {\cal A} (U) \}$,
where for each open subset $U$ in the (globally hyperbolic)
spacetime $(M,g)$, ${\cal A} (U)$
denotes the field operators localized in $U$. Now consider a specific
example, e.g. the quantum theory of a linear Klein-Gordon field on a
globally hyperbolic spacetime $(M,g)$. It is not difficult to see
that if we fix a global Cauchy surface $\Sigma$ in $(M,g)$, and consider
all Lorentz metrics $h$ on $M$ for which $\Sigma$ is a Cauchy surface
for $(M,h)$, then the field algebras ${\cal A}$ corresponding to $(M,
h)$ are all isomorphic to the original algebra ${\cal A}$ of $(M,g)$.
The subalgebras ${\cal A} (U)$, on the other hand, are not mapped onto
each other by the corresponding isomorphisms; i.e. the isomorphisms do
not preserve the subalgebras. This means that the algebra ${\cal A}$ and
the net structure of the subalgebras $\{ {\cal A} (U) \}$ contain
information about the topology of $M$, whereas exactly which subalgebras
of ${\cal A}$ the ${\cal A}(U)$ correspond to contain the geometric
information about the Lorentz metric on $M$. This observation suggests
that the structure consisting of the pair $[{\cal A}, \{ {\cal A}(U) \}
]$ is the right analogue of the observable ${\cal F}$ in quantum theory.

The difficulty now is that ordinarily
we know how to construct local field algebras
only if the background is a globally hyperbolic spacetime, and then only
for linear (noninteracting) fields (with the exception of some
two-dimensional examples). We need a generalization of the algebraic
approach that would make it possible to make sense of quantum field theory
on arbitrary background topological spaces, not just on Lorentzian
manifolds. In the following we will carry out such a generalization;
our general notion of quantum field is abstracted from the usual
algebraic notion by keeping only its most elementary, bare-bones
essentials.

{\bf \noindent 3. A generalization of the algebraic framework for quantum
field theory}

Let $X$ be a topological space. A ``quantum field theory" on $X$ consists
of an abstract $C^{\ast}$ algebra ${\cal A}$ (with identity element $
1$), and a map (which we will
also denote by ${\cal A}$) that associates to each open subset $U$ in
$X$ a closed subalgebra ${\cal A} (U) \subset {\cal A}$ such that the
following two conditions hold:

\noindent (QF1): For every open subset $U \subset X$ $\; {\cal A} (U)$
is a central $C^{\ast}$ algebra, and ${\cal A}
(\{ \}) = {\bf C} \cdot 1$, ${\cal A} (X) = {\cal A}$.

\noindent (QF2): For any collection $\{ W_{\alpha} \}$ of open subsets,
\begin{equation}
{\cal A} ( \bigcup_{\alpha} W_{\alpha} ) = \overline{ <
\bigcup_{\alpha} {\cal A} (W_{\alpha}) > } \; .
\end{equation}
Here $\{ \}$ denotes the empty set, ${\bf C } \cdot 1$
is the $C^{\ast}$ algebra (isomorphic to the algebra $\bf C$ of complex
numbers) generated by $1$,
a {\it central} algebra ${\cal B}$ is one with
the property that its center, \[ Z({\cal B}) \equiv
\{ x \in {\cal B} \, | \, xy=yx \; \forall y
\in {\cal B} \} \; , \] is equal to
${\bf C} \cdot 1 \, $ [$Z({\cal B}) \cong {\bf C}$],
$\; <S>$ denotes the
subalgebra generated by a subset $S \subset
{\cal A}$, and overbar denotes closure in ${\cal A}$.
We will call the theory ${\cal A}$ ``nondegenerate" if ${\cal A}(U)$ is
strictly bigger than ${\bf C}$ for every nonempty open subset $U \subset
X$. Property QF2 implies (but is stronger than)
the well-known ``net-structure" on the
subalgebras $\{ {\cal A} (U) \}$: whenever $U
\subset V$, it holds that ${\cal A}(U) \subset {\cal A}(V)$.
Note that, in standard quantum
field theory, property QF2 does
not in general hold for the local algebra of all ``observables," but it
does hold when ${\cal A}$ consists only of (exponentiated)
smeared field operators.

Now we can incorporate the notion of ``locality" into our general
formulation of field
theory. For this, let for each point $p \in X \, $ $C(p)$ denote the set
\begin{eqnarray}
C(p) & \equiv & \{ q \in X | \not{\exists} \; \mbox{open sets} \; U , \; V \;
\mbox{such that} \nonumber \\
&  & p \in U , \; q \in V  , \; \mbox{and} \;
[{\cal A} (U), {\cal A} (V)] = 0  \} \; ,
\end{eqnarray}
where for $A, \; B \subset {\cal A} $, $\, [A,B]$ denotes the commutator
subalgebra generated by elements of the form $\{ ab-ba \, | \, a \in A,
\; b \in B \}$. The set $C(p)$ consists of those points $q \in X$ that
can ``causally communicate" with $p$
through fields in ${\cal A}$. [We restrict ourselves throughout to bosonic
fields; hence our use of the commutator $[ \; , \; ]$. It is straightforward to
formulate a fermionic version of our discussion by replacing commutators
with anti-commutators. But note that, in the fermionic case, the generators
of the field algebra correspond to the smeared field operators themselves
(which are already bounded) in contrast to the bosonic case, where they
correspond to the exponentiated smeared fields ([5], [6]).]
Some immediately obvious
properties of $C(p)$ are: $q \in C(p) \; \mbox{iff} \; p \in C(q)$,
$C(p)$ is a closed subset of $X$, and,
when ${\cal A}$ is nondegenerate, $p \in C(p) \; \; \forall p \in X$
(this last result follows from QF1). A continuous curve $\gamma :
{\bf R} \longrightarrow X$ is called a ``connector" if
for every $t_{0} \in {\bf R}$ there exists an $\epsilon > 0$ such that
$\gamma (t) \in C[ \gamma (s)] $ for all $t$, $s$ in the interval
$(t_{0} - \epsilon \, , \, t_{0}+ \epsilon )$.
(Thus defined a connector is analogous to a causal curve in spacetime.)
The notion of locality for a quantum field theory ${\cal
A}$ on $X$ is now defined in terms of the topological properties of the
sets $C(p)$. Thus, we will say that ${\cal A}$ is ``weakly local" if the
following two conditions are satisfied:

\noindent (L): There exists an open
neighborhood $V$ around every point $p \in X$
such that for every open neighborhood
$U$ of $p$ contained in $V$ the set $U \cap [C(p) \backslash \{ p \}]$ is
disconnected (here $\backslash$ denotes set difference).

\noindent (WL): For all $p \in X$ $\, C(p)$ is connected.

\noindent The theory $\cal A$ is ``strongly local" if it satisfies
condition L and the following stronger version of WL:

\noindent (SL): For every $p$, $q \in X$ such that $q \in C(p)$ there
exists a connector $\gamma$ joining $p$ and $q$; in particular, the set
$C(p)$ is arcwise connected for all $p \in X$.

\noindent It is easy to see that if condition
L is satisfied ${\cal A}$ must be
nondegenerate. Throughout the rest of this paper we will assume that all
field theories we discuss are nondegenerate unless noted otherwise.

In physical terms, locality provides
for the existence of dynamics, the
``finite speed of propagation" of causal signals.
Accordingly, condition WL (or SL) guarantees that
causal influences propagate from $p$
continuously, and condition L guarantees that signals that communicate
with $p$ propagate with ``finite speed," and that they connect $p$ to disjoint
components of $C(p) \backslash \{ p \}$ (which is necessary if dynamics
at $p$ is to be determined not only by local evolution equations but also
by boundary conditions).

There is a natural notion of ``isomorphism" between quantum field
theories. Let $X$ be a topological space, and let ${\cal A}_{1}$ and ${\cal
A}_{2}$ be field theories on $X$. Then, ${\cal A}_{1}$ and ${\cal A}_{2}$
are said to be isomorphic, denoted
$({\cal A}_{1}, X) \cong ({\cal A}_{2},X)$,
if there is a (isometric) $C^{\ast}$ isomorphism
$\Psi : {\cal A}_{1} \longrightarrow {\cal A}_{2}$ such that for every
open subset $U \subset X$, \[ {\cal A}_{2} (U) = \Psi [ {\cal A}_{1} (U)
] \; . \] As an example, let ${\cal A}$ be an ordinary
(e.g. Klein-Gordon) field theory on a globally hyperbolic spacetime
$(M,g)$, $\, h : M \longrightarrow M$ be a diffeomorphism of $M$, and
define a new theory $h^{\ast} {\cal A}$ on $M$ by (for
all open $U \subset M$) \[
(h^{\ast} {\cal A} ) \: (U) \equiv \left. {\cal A}^{(KG,m)} \right|_{(M,
h^{\ast}g)}  [h^{-1} (U)] \]
[where the field theory that
appears on the right-hand
side is the standard Klein-Gordon theory corresponding to the spacetime $(M,
h^{\ast}g)$]. It is then easy to see that the theories ${\cal A}$ and
$h^{\ast} {\cal A}$ thus defined are isomorphic over $M$.
Hence our notion of isomorphism is a natural
generalization of the usual diffeomorphism invariance in curved-spacetime
field theory.

How good a job do the sets $C(p)$ ($p \in M$) do in
providing a generalized ``causal
structure" on the space $X$? A partial answer is given by the following
result:

{\noindent \it Theorem}: Let $X$ be a locally compact topological space,
and $\cal A$ a quantum field theory on $X$. For $U$, $V
\subset X$ open subsets, the subalgebras
${\cal A} (U)$ and ${\cal A} (V)$ {\it
fail} to commute if and only if
there exists a point $p \in U$ and a point $q \in
V$ such that $q \in C(p)$.

{\noindent \it Proof}: We first prove the following:

{\noindent \it Lemma}: Let $X$ be a topological space and ${\cal A}$ a
quantum field theory on $X$. If $p \in X$ and $U \subset X$
open are such that $\overline{U}$ is compact and
$C(p) \cap \overline{U} = \{ \}$, then there exists an open neighborhood
$V$ of $p$ such that ${\cal A}(U)$ and ${\cal A}(V)$ commute.

{\noindent \it Proof of the Lemma}: Since $\overline{U} \cap C(p) = \{
\} $, $\forall q \in \overline{U}$ there exists an open neighborhood $W_{q}$
of $q$ and an open neighborhood
$M_{q}$ of $p$ such that ${\cal A} (W_{q})$ and ${\cal
A}(M_{q})$ commute. Now $\{ W_{q} | q \in \overline{U} \}$ is a covering
of $\overline{U}$ and $\overline{U}$ is compact; this implies that there
exists a finite set of points $q_{1}, \cdots , q_{n} \in \overline{U}$
such that $W_{q_{1}} \cup \cdots \cup W_{q_{n}} \supset \overline{U}
\supset U$. But since ${\cal A} (W_{q_{1}} \cup \cdots \cup W_{q_{n}})
= \overline{< \bigcup_{i=1}^{n} {\cal A}(W_{q_{i}}) >}$, and
${\cal A} (M_{q_{1}} \cap \cdots \cap M_{q_{n}}) \subset
\bigcap_{i=1}^{n} {\cal A}(M_{q_{i}})$, $\; {\cal A}(W_{q_{1}} \cup \cdots
\cup W_{q_{n}})$ commutes with ${\cal A}(M_{q_{1}} \cap \cdots \cap
M_{q_{n}})$. Thus ${\cal A}(U) \subset {\cal A}(W_{q_{1}} \cup \cdots
\cup W_{q_{n}})$ commutes with ${\cal A} (M_{q_{1}} \cap \cdots \cap
M_{q_{n}})$, $\;$and $V \equiv M_{q_{1}} \cap \cdots \cap M_{q_{n}}$ is an
open neighborhood of $p$ satisfying the desired condition. This
completes the proof of the lemma.

\noindent Now back to the proof of the theorem. If there exist points $p
\in U$ and $q \in V$ such that $q \in C(p)$, then by the definition of
$C(p)$ we have that ${\cal A}(U)$ and ${\cal A}(V)$ must fail to
commute. To prove the implication in the converse direction,
it clearly suffices
to show that if $V
\cap \bigcup_{p \in U} C(p) = \{ \}$ then ${\cal A} (U)$ and ${\cal A} (V)$
commute. Let $V \cap \bigcup_{p \in U} C(p) = \{ \} $. Since $X$ is locally
compact, every $q \in V$ has an open neighborhood $S_{q}$ of compact
closure such that $\overline{S_{q}} \subset V$. By the lemma, for each
fixed $q \in V$ and given $p \in U$ there
exists an open neighborhood $W_{qp}$ of $p$ such that ${\cal A} (W_{qp})$
and ${\cal A} (S_{q})$ commute. This implies that ${\cal A}(S_{q})$ and
${\cal A}(\bigcup_{p \in U} W_{qp})$ commute, which implies ${\cal A}
(S_{q})$ commutes with ${\cal A}(U)$. Since this is true for every $q
\in V$ (and the neighborhoods $S_{q}$ cover $V$), we conclude that ${\cal
A}(V)$ commutes with ${\cal A}(U)$. $\Box$

Another fundamental causal notion that can be naturally formulated
within our general framework is that of ``domain of dependence." Namely,
an open subset $U \subset X$ is in the domain of dependence of another
open set $V \subset X$ if ${\cal A}(U)$ is contained in ${\cal A}(V)$.
Physically, this corresponds to a situation where all fields localized
in $U$ can be obtained by evolving fields in $V$ via the
local dynamics. To formalize this idea more elegantly, we introduce the
following:

{\noindent \it Definition}: An open subset $U \subset X$ is called a
``diamond" if for all open $V \subset X$ $\; {\cal A}(V) \subset {\cal
A}(U)$ implies $V \subset U$.

\noindent Diamonds enjoy a number of properties all of
which follow readily from
the above definition. Namely: (in the following $U$, $V$, $W_{\alpha}$
denote open sets in $X$)

\noindent (i): The space $X$ is a diamond, and the empty set $\{ \}$
is a diamond if and only if ${\cal A}$ is nondegenerate.

\noindent (ii): If $W_{\alpha}$ are diamonds, then $\mbox{Int}(
\bigcap_{\alpha}W_{\alpha})$ is a diamond. Here $\mbox{Int}(A)$ denotes
the topological interior of $A$.

{\noindent \it Theorem}: $\forall $ open $U \subset X$ there exists a
smallest diamond $D(U)$ containing $U$; more precisely, there exists a
diamond $D(U)$ such that $D(U) \supset U$ and if $V$ is any other
diamond containing $U$ then $V \supset D(U)$.

{\noindent \it Proof}: $D(U)$ is uniquely given by $\mbox{Int}[
\bigcap_{\alpha}(D_{\alpha})]$, where the intersection is over all
diamonds $D_{\alpha}$ containing $U$.
$\Box$

\noindent (iii): $U$ is a diamond if and only if $U = D(U)$. Hence
$\forall U$ $\; D(D(U))=D(U)$.

\noindent (iv): $D(U) = \bigcup_{\alpha} W_{\alpha}$, where the union is
over all open $W_{\alpha}$ such that ${\cal A}(W_{\alpha}) \subset {\cal
A} (U)$. As a consequence, ${\cal A}(D(U)) = {\cal A}(U)$ $\; \forall U$.

\noindent (v): If $V$ is a diamond and $V \supset U$, then $V \supset
D(U)$.
{}From this it follows that
${\cal A}(U) \subset {\cal A}(V)$ implies $D(U) \subset
D(V)$. In particular, ${\cal A}(U) = {\cal A}(V)$ if and only if $D(U) =
D(V)$.

\noindent (vi): $D(U \cap V) \subset D(U) \cap D(V)$; equality does not
hold in general.

\noindent (vii): $D(\bigcup_{\alpha}W_{\alpha}) = D[\bigcup_{\alpha}
D(W_{\alpha})]$.

\noindent For an open subset $U \subset X$, the
domain of dependence of $U$ is the largest open set in which all local
fields are dynamically determined by the fields localized in $U$. It is
clear from the above properties that the diamond
$D(U)$ is precisely the domain of
dependence.

We started this discussion by pointing out that in quantum field theory
the structure consisting of
${\cal A}$ and the subalgebras $\{ {\cal A}(U) \}$
is the fundamental observable associated with spacetime topology, the
sought-after analogue
in quantum theory of the observable ${\cal F}$
of classical physics.
But we have not yet explained how the topological space $(X, \tau )$
can be recovered from a knowledge of $[{\cal A}, \{ {\cal A}(U) \}]$.
Indeed, this task now appears quite nontrivial: it
seems impossible to distinguish,
solely by observing the subalgebras ${\cal A}(U)$,
the open subsets $U \subset X$ from their
associated diamonds $D(U)$,
since, as we have seen above, ${\cal A}(U) = {\cal A}(D(U))$ for
every open set. The true observable, then, is the collection $\{ {\cal A} (D)
\}$, where $D$ ranges over all diamonds in $X$. How can we reconstruct $(X,
\tau )$ using only the structure of its diamonds?

We will answer this question shortly. First, however, we will study
the essential structure in the observables $\{ {\cal A} (U) \}$
and $\{ {\cal A}(D) \}$ which contains the information relevant to
spacetime topology; namely, the partial order on the sets $\{ {\cal A}
(U) \}$ and $\{ {\cal A}(D) \}$, given by open-set inclusion, $U
\subset U^{\prime}$, in the first case, and diamond inclusion, $D \subset
D^{\prime}$, in the second [by (v) above, the partial order
on $\{ {\cal A}(D) \}$ in this second case coincides with
that given by subalgebra inclusion]. More
precisely, the structure of $\{ {\cal A}(U) \}$ is that of a
``$C^{\ast}$ frame," and the structure of $\{ {\cal A}(D) \}$ is that
of a ``$C^{\ast}$ lattice." This leads us naturally into our final,
ultimate generalization of quantum field theory: The structure of the
observables $\{ {\cal A} (U) \}$ [or of $\{ {\cal A}(D) \}$] makes no
explicit reference to the background topological space $(X, \tau )$.
In quantum field theory, the
fundamental information about spacetime structure
is contained in the partial ordering of the set $\{ {\cal A}(U)
\}$, and, therefore, it is natural to consider the category of such
ordered-subalgebras as the proper domain to which spacetime structure
ultimately belongs in quantum theory.

Let us, then, proceed to a general discussion of this new category.

{\bf \noindent 4. Quantum lattices and quantum frames}

We assume throughout this section that the reader is familiar with the
basic notions of lattice theory, including frames and their connection
with topological spaces. For a systematic treatment of these subjects,
see [8] and [9]; alternatively, the Appendix below presents a quick
review of the relevant concepts. The
background contained in the Appendix is sufficient for following
our discussion in this section.

A $C^{\ast}$ lattice is a complete lattice $L$ which, as a set,
consists of closed
subalgebras of an abstract $C^{\ast}$ algebra ${\cal A}$ (with
identity),
and where the partial order on $L$, given by the usual
subalgebra inclusion, satisfies the following properties: (i)
the least element ${\bf 0}$ is ${\bf C}$ and
the largest element ${\bf 1}$ is ${\cal A}$, and (ii) the join
$\bigvee_{\alpha} a_{\alpha}$ (for $a_{\alpha} \in L$) is given by
$\overline{<\bigcup_{\alpha}a_{\alpha}>}$. A simple example of a
$C^{\ast}$ lattice is the complete lattice $M$ of {\it all} closed
subalgebras of a $C^{\ast}$ algebra ${\cal A}$ (with identity). In $M$,
the meet operation is simply $\bigwedge_{\alpha}a_{\alpha} =
\bigcap_{\alpha}a_{\alpha}$. Note that
for a more general $C^{\ast}$ lattice $L$ based on
${\cal A}$ the meet in $L$ would not, in general, have this simple form;
in other words, $L$ is not necessarily a sublattice of $M$ (see
Appendix), even though the partial order on $L$ is induced from that on
$M$. More nontrivial examples of $C^{\ast}$ lattices are those given by
$\{ {\cal A} (D) \}$, where ${\cal A}$ is a quantum field theory on a
topological space $X$ and $D$ ranges over all diamonds in $X$ (see
Sect.\,3).

A $C^{\ast}$ frame $(F, {\cal A})$ consists of a frame $F$,
an abstract $C^{\ast}$ algebra ${\cal A}$ (with identity),
and a map ${\cal A}$ that
associates to each $b \in F$ a closed subalgebra
${\cal A}(b) \subset {\cal A}$ such that the conditions
\begin{equation}
{\cal A}({\bf 0}) = {\bf C} \; , \; \; \; \; \; \;
{\cal A}({\bf 1}) = {\cal A} \; ,
\end{equation}
and
\begin{equation}
{\cal A}(\bigvee_{\alpha} b_{\alpha} ) = \overline{<
\bigcup_{\alpha} {\cal A}(b_{\alpha}) >} \;, \; \; \; \; b_{\alpha} \in F
\;
\end{equation}
are satisfied. Typical examples of quantum frames are, of course, those
in which $F$ is the open-set frame
$\Omega (X)$ of a topological space $X$, and the
map ${\cal A}$ associates to each open set $b \in F$ the local field algebra
${\cal A}(b)$ of a quantum field theory ${\cal A}$ on $X$. Recall (see
Appendix) that given a frame $F$, we have a canonical construction which
associates to $F$ a topological space $\mbox{pt}(F)$. As long as the
space $X$ is ``reflexive" (see Appendix), it can be reconstructed
from its open-set frame $\Omega (X)$ as $\mbox{pt}[ \Omega (X) ]$. For
example, all Hausdorff $X$ are reflexive. Therefore, provided spacetime is
at least Hausdorff, its topology can be completely recovered from the
observable $[{\cal A}, \{ {\cal A}(U) \}]$ simply by using the frame
structure of the $C^{\ast}$ frame $\{ {\cal A}(U) \}$. In other words, for a
reflexive $X$, a quantum field theory ${\cal A}$ on $X$ contains
precisely the same information as the open-set
frame $\Omega (X)$, endowed with the $C^{\ast}$-frame structure
given by the theory ${\cal A}$.
This is the
reconstruction we promised at the end of the last section; it is the
quantum analogue of the reconstruction of a Tychonoff $X$ from the classical
observable ${\cal F}$ (see Sect.\,2). On the other hand,
the natural category to which the notion of ``spacetime structure" belongs
now becomes the category of frames and frame maps,
manifestly larger than
the category of topological spaces and continuous
functions. Indeed, there is no reason to assume a priori that
$F$ for a $C^{\ast}$ frame $(F, {\cal A})$ is the open-set frame of
{\it any} topological space; in general it is not, as $F$ coincides with
$\Omega [ \mbox{pt} (F) ]$ only for a restricted class of frames
(called ``spatial" frames; see Appendix). We will discuss
some of the implications of this generalization
in the next paper of this series (see the concluding section below).
In the remainder of this paper, we will briefly explore some elementary
properties of quantum lattices and frames; but first, we will pursue
the answer to the second question posed at the end of the last section:
namely the reconstruction of $X$ from the observable $\{ {\cal A}(D) \}$
(where $D$ ranges over the diamonds in $X$).

We have seen that $\{ {\cal A}(D) \}$ has the structure of a
$C^{\ast}$ lattice, and that there is a canonical construction which
associates to each frame $F$ a topological space $\mbox{pt}(F)$. The
reconstruction we desire would be described (for sufficiently regular
spaces $X$) once we describe how
to construct a $C^{\ast}$ frame $(F, {\cal A})$
canonically
associated to every $C^{\ast}$ lattice $L$ (based on
${\cal A}$). We will now present such a
construction:

Let $L$ be {\it any} complete lattice. We first describe how to
construct a
frame $F(L)$ canonically associated to $L$. The main idea here is to see
the lattice $L$ as a ``tiling" of the frame $F(L)$. Think of the
following concrete example as a model: Let $F$ be the open-set frame
of the Euclidean space ${\bf R}^{2}$, and let $L$ be the lattice of all
open rectangles, i.e. open sets in ${\bf R}^{2}$ of the form $I \times
J$, where $I$ and $J$ are open intervals in ${\bf R}$ (partial order on
$L$ is the usual inclusion, and the meet is $\bigwedge_{\alpha}U_{\alpha}
= \mbox{Int}(\bigcap_{\alpha}U_{\alpha})$; the same as the meet
operation in the open-set frame $F$).
In this example,
there is a clear intuitive sense in which $L$ ``tiles" $F$, and it seems
obvious that $F$ should be constructible entirely in terms of $L$. Our
construction is an abstract scheme in which this idea is made precise.

So given an arbitrary complete lattice $L$ with join $\vee$ and meet
$\wedge$, let $2^{L}$ denote the set
of all subsets of $L$, and introduce an equivalence relation $\sim$ on
$2^{L}$ by
\begin{equation}
\mbox{for} \; K, \; K^{\prime} \subset L \; , \; \;
K \sim K^{\prime} \Longleftrightarrow \bigvee_{k \in K} x \wedge k
= \bigvee_{k^{\prime} \in K^{\prime}} x \wedge k^{\prime} \; \;
\; \forall x \in L \; .
\end{equation}
There is a natural partial order on the quotient set $2^{L} \, / \sim $,
namely
\begin{equation}
[K] \leq [K^{\prime}] \Longleftrightarrow
\bigvee_{k \in K} x \wedge k \leq \bigvee_{k^{\prime} \in K^{\prime}}
x \wedge k^{\prime} \; \; \; \forall x \in L \; ,
\end{equation}
where $[K]$ denotes the equivalence class in $2^{L} \, / \sim$ of the subset
$K \in 2^{L}$, and
the order $\leq$ on the right hand side is that of the original
lattice $L$. The definition (7) is clearly independent of which
representatives $K$, $K^{\prime}$ are chosen for the classes $[K]$ and
$[K^{\prime}]$ [see Eq.\,(6)]. Now put $F(L) \equiv 2^{L} \, / \sim$.
It is easy to verify that the set
$F(L)$ under the partial order (7) is a complete
lattice, and that the join ${\vee}^{F}$ and meet ${\wedge}^{F}$ of
$F(L)$ are simply $[K] {\vee}^F [K^{\prime}] = [K \cup K^{\prime}]$
and $[K] {\wedge}^{F} [K^{\prime}] = [K \wedge K^{\prime}]$, where $K
\wedge K^{\prime}$ denotes the subset $K \wedge K^{\prime} \equiv \{
k \wedge k^{\prime} \, | \, k \in K, \; k^{\prime} \in K^{\prime} \}$.
In fact we
have, more generally,
\begin{equation}
{\bigvee_{\alpha}}^{F} [K_{\alpha}] = [\bigcup_{\alpha}K_{\alpha}]
\; , \; \; \; \; \; \; {\bigwedge_{\alpha}}^{F} [K_{\alpha}] = [
\bigwedge_{\alpha} K _{\alpha} ] \; ,
\end{equation}
where $[K_{\alpha}]$ are an arbitrary collection of elements of $F(L)$. [It
is a pleasant exercise to verify that the definitions (8)
do not depend on which representatives $K_{\alpha}$ are chosen for the
classes $[K_{\alpha}] \in F(L)$.] It is also obvious that the least
element ${\bf 0}$ of $F(L)$ is $[ \{ \} ]$ and the largest element $\bf 1$
is $[L]$. We now claim that $F(L)$ with this complete
lattice structure is, in fact, a frame, i.e. satisfies the join-infinite
distributive identity [Appendix, Eq.\,(12)].
To see this, we simply observe
that for $[M], \; [K_{\alpha}] \in F(L)$,
\begin{eqnarray}
[M] \, {\wedge}^{F} \; {\bigvee_{\alpha}}^{F} [K_{\alpha}] & = &
[M \wedge \bigcup_{\alpha} K_{\alpha} ] = [ \bigcup_{\alpha}  M \wedge
K_{\alpha} ] \nonumber \\
& = & {\bigvee_{\alpha}}^{F} [M \wedge K_{\alpha}] =
{\bigvee_{\alpha}}^{F} [M] \, {\wedge}^{F} \; [K_{\alpha}] \; .
\end{eqnarray}
There exists a natural imbedding $i : L \longrightarrow F(L)$ given by
$i: k \in L \mapsto [\{ k \}] \in F(L)$. It is not difficult to show that
$i$ is one-to-one and order preserving in both directions, i.e. $k \leq
k^{\prime} \Longleftrightarrow i(k) \leq i(k^{\prime})$. Thus $L$ is
imbedded in $F(L)$ in such a way that the order on $L$ coincides with
that induced from $F(L)$, and the meet $\wedge$ of $L$ coincides
with the meet ${\wedge}^{F}$ of $F(L)$. Moreover, $F(L)$ is generated by
$L$ under this imbedding since, clearly, $\forall \; [K] \in F(L)$ we have
$[K] = [\bigcup_{k \in K} k] = {{\bigvee}^{F}}_{k \in K} [ \{ k \} ] $.
Hence the lattice $L$ is imbedded in $F(L)$ as a ``tiling," just like the
example we described in the previous paragraph. Note, also, that if $L$
itself is a frame, then $F(L)$ simply coincides with $L$ [see
Eqs.\,(6)$-$(7)]. This completes
our construction of the frame $F(L)$ associated to an arbitrary lattice
$L$.

Now let $L$ be a $C^{\ast}$ lattice based on a $C^{\ast}$ algebra ${\cal
A}$. Construct the frame $F(L)$
associated to $L$ as in the above paragraph. Each element $b$ of
$F(L)$ is an equivalence class $[A]$ of subsets of $L$. Moreover, by
Eq.\,(6), for any two representatives $A$, $A^{\prime}$ of $[A]$ we have
$\bigvee_{a \in A} a = \bigvee_{a^{\prime} \in A^{\prime}} a^{\prime}$
[just take $x = {\bf 1}$ in Eq.\,(6)]. For a $C^{\ast}$ lattice $L$,
where the join is simply the closure of the subalgebra generated by its
arguments, this implies that $\overline{<\bigcup_{a \in A} a >}$ is a
well defined $C^{\ast}$ algebra for each $b = [A]$ in $F(L)$. The
assignment of this algebra to each element $b$ of the frame $F(L)$ gives
$F(L)$ the structure of a $C^{\ast}$ frame [see Eqs.\,(7) and (8)]. This
is the canonical $C^{\ast}$ frame $[F(L), {\cal A}]$ associated to the
$C^{\ast}$ lattice $L$.

It is also possible to go back from a $C^{\ast}$ frame $(F, {\cal A})$ to a
$C^{\ast}$ lattice $L(F)$ of ``diamonds" of $F$, where, in the context of
a $C^{\ast}$ lattice $(F, {\cal A})$,
diamonds are defined analogously to Sect.\,3 as
elements $d \in F$ such that $\forall \, b \in F$
$\; {\cal A}(b) \subset {\cal A} (d)$
implies $b \leq d$.

Finally, we come to the formal definition of ``quantum frames" and
``quantum lattices." A quantum frame is essentially a $C^{\ast}$ frame
$(F, {\cal A})$, with the additional technical conditions that (i) $F$ is
generated by its atoms, (ii) $\forall b \in F$, $b < {\bf 1}$, there
exists a completely prime filter disjoint from $j^{-}(b) \equiv \{ d \in F \,
| \,
d \leq b \}$ (see Appendix for the definitions of these terms), and (iii)
${\cal A}(b)$ is a central $C^{\ast}$ algebra $\forall b \in F$. A
quantum lattice is a $C^{\ast}$ lattice $L$ whose associated
$C^{\ast}$ frame $(F, {\cal A})$ is a quantum frame. These extra
restrictions prove useful in discussing certain physical properties of
quantum frames and lattices, as we will see when we study examples of
these objects in our next paper of the series.

For a foretaste of the kind of physics one can study in this framework,
let us try to define a notion of locality for quantum frames. Let $(F,
{\cal A})$ be a quantum frame, and for any $b \in F$ define the ``causal
complement" $S(b)$ of $b$ as the set
\begin{equation}
S(b) \equiv \{ d \in F \; | \; [{\cal A}(d), {\cal A}(b)] = 0 \; \} \; .
\end{equation}
Define the ``causal trace" of $b$ as the subset
\begin{equation}
J(b) \equiv \{ d \in F \; | \; j^{-}(d) \cap S(b) = \{ {\bf 0} \} \; \}
\; .
\end{equation}
Recall that for a subset $A \subset F$, $\; \bigvee A$ denotes the
element ${\bigvee_{a \in A}}a$. Let for each $b \in F$ $\; \tilde{b}$
denote the ``exterior" of $b$ defined by $\tilde{b} \equiv \bigvee
\{ d \in F |
d \wedge b = {\bf 0} \}$. Now we can define
a quantum frame $(F , {\cal A})$
to be local if it has the following two
properties:

\noindent (i) For every atom $b \in F$ the element $\bigvee J(b)$ is an
atom.

\noindent (ii) Every completely prime filter ${\cal P}$ in $F$ admits a
basis ${\cal B}$ and an element $f \in {\cal P}$ such that for each $e
\in {\cal B}$ the element $f \wedge \tilde{e} \wedge \, \bigvee J(e) $
is composite.

\noindent It is not hard to see that this definition is designed to be
as close as possible within the frame category
to the general notion of locality introduced in Sect.\,3 for topological
spaces and quantum fields. Note, also, that we can define a local
quantum lattice simply as a quantum lattice whose associated quantum
frame is local.

As a last remark, we note that
just as in the case of the general framework of Sect.\,3, so here
also we have a natural generalization of the notion of ``diffeomorphism
invariance" in terms of the notion of isomorphism
between quantum lattices and frames.
An isomorphism between two quantum lattices
(or general $C^{\ast}$ lattices) $L_{1}$ and
$L_{2}$ is defined simply as a $C^{\ast}$-algebra isomorphism $\Psi :
{\cal A}_{1} \longrightarrow {\cal A}_{2}$ which carries the subalgebras
that constitute $L_{1}$ onto those that constitute $L_{2}$. Similarly,
an isomorphism between two quantum frames (or more general $C^{\ast}$
frames) $(F_{1}, {\cal A}_{1})$ and
$(F_{2}, {\cal A}_{2})$ is a pair $(f, \Psi )$, where $f : F_{1}
\longrightarrow F_{2}$ is a frame isomorphism, $\Psi : {\cal A}_{1}
\longrightarrow {\cal A}_{2}$ is a $C^{\ast}$ algebra isomorphism, and
the two are compatible in the sense that \[
\Psi [{\cal A}_{1} (b)] = {\cal A}_{2} [f(b)] \]
for all $b \in F_{1}$.

{\bf \noindent 5. Conclusion}

We have finally obtained a quite promising, if not entirely satisfactory,
answer to the original question posed in the Introduction; namely, we
now have a rather precise understanding of the physical origin of
spacetime topology, or, more generally, of spacetime structure. The most
promising aspect of our approach so far is the fact that it naturally
suggests a candidate for the fundamental, ``discrete" structure
of spacetime at arbitrarily small length scales, namely the structure of
a quantum lattice or frame.
There are a number of advantages to this suggestion for the discrete
structure of spacetime: For one, our formalism does not treat
spacetime as a set of
``points" equipped
with some (topological, causal, ...) preferred structure. Points of a frame
$F$ do not have fundamental physical reality, even though they
may be constructed abstractly
as ``completely prime filters" in the lattice structure of $F$. This is
in agreement with the intuitive expectation that in full quantum gravity
``topology fluctuations" would prevent one from defining spacetime points in
a sensible way. Also, in our approach the notion of ``diffeomorphism
invariance" finds a very natural and simple reformulation in terms of
isomorphisms of quantum frames. And, finally,
quantum frames have a quantized structure at the outset; no extra
``quantization" step is necessary to study their implications for the
small-scale structure of spacetime.

Of course, much remains to be done before we can determine whether our
approach can ever contribute to the
larger goal of gaining new insights into quantum gravity. Here we
focused our attention mainly on developing the mathematical foundations
for our viewpoint; hence our discussion has been rather general and
abstract. In the next paper ([10]) of this series we will tackle some of the
more ``physical" questions raised by our discussion above. Among the
subjects of this forthcoming manuscript
are a discussion of quantum lattices generated by the usual quantum
field theory models in curved spacetime, the analysis of some simple
toy-models for ``discrete" quantum frames and lattices, and a discussion
on the role and possible quantization of the ``metric" in our approach.

{\bf \noindent Appendix: Some basic facts about lattices and frames}

What we will give here is mainly an explanation of the terminology used and
the statements (without proofs)
of the results referred to in the discussion of Sect.\,4. We
refer the reader to the sources [8] and [9] for
proofs and for more detailed information.

A {\bf lattice} is a set $L$ with a partial order $\leq$ such that for
every pair of elements $a, \; b \in L$ the (unique) least upper bound,
$\mbox{l.u.b.}(a,b)$, and the greatest lower bound, $\mbox{g.l.b}(a,b)$,
both exist. These elements are denoted by $a \vee b$, and $a \wedge b$,
respectively, and both $\vee$ (called the ``join") and $\wedge$ (called
the ``meet") are commutative, associative binary operations on $L$. A
{\bf complete lattice} is a lattice in which the meet and join of
arbitrary (not just finite) collections of elements exist. A {\bf frame}
is a complete lattice $F$ in which the {\bf join-infinite distributive
identity}
\begin{equation}
a \wedge \bigvee_{\alpha}b_{\alpha} = \bigvee_{\alpha} a \wedge
b_{\alpha} \; , \; \; \; \; \forall \, a, \; b_{\alpha} \in F \;
\end{equation}
holds. If $A \subset L$ is a subset of a complete lattice $L$,
we denote by $\bigvee A$ and $\bigwedge A$ the elements
$\bigvee_{a \in A} a$ and $\bigwedge_{a \in A} a$ of $L$.
In any complete lattice $L$, there is defined a least element
${\bf 0} \equiv \bigwedge L$, and a largest element ${\bf 1} \equiv
\bigvee L$.
An example of a complete lattice is the lattice of all subsets of a set
$A$ partially ordered by inclusion. In this case, $\vee$ and $\wedge$
coincide with the set union $\cup$ and set intersection $\cap$,
respectively. This lattice is always a frame. For an example of a
complete lattice which is not a frame, consider the lattice of all open
rectangles in ${\bf R}^{2}$ discussed in Sect.\,4. Other
typical examples of frames are given by the open-set lattices, denoted
$\Omega (X)$, of topological spaces $(X, \tau )$.
In $\Omega (X)$, the join is the usual set union, but the meet is given
by $\bigwedge_{\alpha} U_{\alpha} = \mbox{Int}
(\bigcap_{\alpha}U_{\alpha})$, which generally coincides with set intersection
only for finite collections $\{ U_{\alpha} \}$ of open sets.

Let $L_{1}$, $L_{2}$ be lattices, and $f : L_{1} \longrightarrow L_{2}$
be a map. Then, denoting by $a, \; a_{\alpha}$ and $b$ arbitrary elements
of $L_{1}$, we can introduce the following notions:

\noindent (i) $f$ is called {\bf order preserving} if $a \leq b
\Longrightarrow f(a) \leq f(b)$.

\noindent (ii) $f$ is {\bf order preserving in both directions} if $a
\leq b \Longleftrightarrow f(a) \leq f(b)$.

\noindent (iii) $f$ is a {\bf lattice homomorphism} if $f(a \vee b) = f(a)
\vee f(b)$ and $f(a \wedge b) = f(a) \wedge f(b)$. If $L_{1}$ and
$L_{2}$ are complete, then $f$ is a {\bf complete lattice homomorphism}
if $f(\bigvee_{\alpha} a_{\alpha}) = \bigvee_{\alpha}f(a_{\alpha})$ and
$f(\bigwedge_{\alpha}a_{\alpha}) = \bigwedge_{\alpha}f(a_{\alpha})$.

\noindent (iv) $f$ is a {\bf lattice isomorphism} if it is one-to-one,
onto, and a lattice homomorphism; or, equivalently, if it is onto and order
preserving in both directions.

\noindent (v) If $L_{1}$ and $L_{2}$ are frames, then $f$ is a {\bf
frame map} if it is a lattice homomorphism that preserves arbitrary
joins and maps ${\bf 0} \mapsto {\bf 0}$ and ${\bf 1} \mapsto {\bf 1}$.
$f$ is a {\bf frame isomorphism} if it is both a frame map and a lattice
isomorphism.

\noindent Notice that frame maps generalize inverses of continuous functions;
i.e., if $X$, $Y$ are topological spaces and $f: X \longrightarrow Y$ is
a continuous map, then $f^{-1} : \Omega (Y) \longrightarrow \Omega (X)$
is a frame map.
Hence the category of frames and frame maps is a generalization of the
category of topological spaces and continuos functions.

In any lattice $L$, for $a \in L$ $\; j^{-}(a)$ denotes the subset
$j^{-}(a) = \{ b \in L \, | \, b \leq a \}$, and, more generally, for a
subset $A \subset L$ $\; j^{-} (A)$ is defined as $j^{-}(A) \equiv \{ b
\in L \, | \, \exists a \in A \; \mbox{such that} \; b \leq a \}$.
$j^{+}(a)$ and $j^{+}(A)$ are defined similarly. An element $a \in L$ is
called an {\bf atom} if $a$ cannot be written in the form $b \vee c$
with $b \wedge c = {\bf 0}$; if $a$ {\it can} be written in this form it
is called {\bf composite}.

Let $\{ 0, 1 \}$ denote the unique frame with two elements. If $F$ is
any frame, we define
\begin{equation}
\mbox{pt} (F) \equiv \{ p : F \longrightarrow \{ 0, 1 \} \; | \;
p \; \mbox{is a frame map} \, \} \;.
\end{equation}
The elements of $\mbox{pt}(F)$ are called the {\bf points} of $F$.
Now consider all subsets of the set $\mbox{pt}(F)$ which have the form
${\cal O}_{b} \equiv \{ p \in \mbox{pt}(F) \, | \, p(b) = 1 \}$ for some
$b \in F$. It is not hard to show that the collection of subsets $\{
{\cal O}_{b} \subset \mbox{pt}(F) \, | \, b \in F \}$ defines a topology
on $\mbox{pt}(F)$. By $\mbox{pt}(F)$ we will always denote this
canonical topological space associated to the frame $F$. It follows that
$\mbox{pt}(F)$ is always $T_{0}$. A frame $F$ is called {\bf spatial} if
$F$ is frame-isomorphic to $\Omega[\mbox{pt}(F)]$. A topological space
$X$ is called {\bf reflexive} if $X$ is homeomorphic to
$\mbox{pt}[\Omega (X) ]$. It can be shown that all Hausdorff spaces are
reflexive.

Let $L$ be an arbitrary lattice. A subset $S \subset L$ is called a {\bf
sublattice} if $S$ is closed under the operations $\wedge$ and $\vee$ of
$L$. A subset $P \subset L$ is called a {\bf
filter} if $a, \; b \in P$ implies $a \wedge b \in P$ and $a \in P, \;
b \geq a$ implies $b \in P$. Equivalently, a filter $P$ is a sublattice
for which $a \in P$ implies $a \vee b \in P$ for all $b \in L$. For any
element $a \in L$, the set $j^{+}(a)$ is a filter.
Let $P \subset L$ be a filter. A {\bf basis} for $P$ is a subset ${\cal
B} \subset L$ such that $a \in P \Longleftrightarrow a \geq b$ for some
$b \in {\cal B}$; in other words, $P = j^{+}({\cal B})$. A filter $P \subset
L$ is called {\bf prime} if for all $a, \; b \in L$ $\; a \vee b \in
P$ implies either $a \in P$ or $b \in P$. For example,
in the lattice of all subsets of a set $A$, filters of the form
$j^{+}( \{ a \} )$ are prime for all $a \in A$.
If $L$ is a complete lattice and $P
\subset A$ is a filter, $P$ is called {\bf completely prime} if
for any arbitrary collection $\{ a_{\alpha} \} \subset L$
$\; \bigvee_{\alpha} a_{\alpha} \in P$ implies that at least one of the
$a_{\alpha}$ is contained in $P$. The filters $j^{+}( \{ a \} )$
in the example we have
just given are completely prime. It can be shown that, for a frame $F$,
a subset ${\cal P} \subset F$ is a completely prime filter if and only
if there exists a frame map $f : F \longrightarrow \{ 0,1 \}$ such that
${\cal P} = f^{-1}(1)$. In other words, the set $\mbox{pt}(F)$ of ``points"
of a frame $F$ is in one-to-one correspondence with the set $\{ {\cal P}
\}$ of all
completely prime
filters in $F$.

\newpage

{\bf \noindent REFERENCES}

\noindent{\bf 1.} R. Penrose, in {\it Quantum Theory and Beyond}, Ted
Bastin ed. (Cambridge University Press, Cambridge 1971).

\noindent{\bf 2.} L. Bombelli, L. Lee, D. Meyer and R. Sorkin, Phys.
Rev. Letters {\bf 59}, 521 (1987); ibid. {\bf 60}, 656 (1987).

\noindent{\bf 3.} C. J. Isham, {\it An Introduction to General Topology
and Quantum Topology}, lectures presented at the Advanced Summer
Institute on Physics, Geometry and Topology, Banff, August 1989.

\noindent{\bf 4.} S. Willard, {\it General Topology} (Addison-Wesley,
Reading, Mass. 1970).

\noindent{\bf 5.} B. S. Kay and R. M. Wald, Phys. Reports {\bf 207}, 49
(1991).

\noindent{\bf 6.} R. M. Wald, in: {\it Quantum Mechanics in Curved
Spacetime}, J. Audretsch and V. de Sabbata eds. (Plenum, New York,
1992).

\noindent{\bf 7.} U. Yurtsever, {\it Algebraic approach to quantum field
theory in non-globally-hyperbolic spacetimes}, UCSB Physics preprint
UCSBTH-92-43,
October 1992.

\noindent{\bf 8.} G. Graetzer, {\it General Lattice Theory} (Birkhauser,
Basel 1978).

\noindent{\bf 9.} P. T. Johnstone, {\it Stone Spaces} (Cambridge
University Press, Cambridge 1982).

\noindent{\bf 10.} U. Yurtsever, {\it Quantum lattices and
quantum frames}, UCSB Physics Preprint (in preparation).

\end{document}